\documentclass[fleqn,usenatbib]{mnras}

\usepackage{newtxtext,newtxmath}

\usepackage[T1]{fontenc}

\DeclareRobustCommand{\VAN}[3]{#2}
\let\VANthebibliography\thebibliography
\def\thebibliography{\DeclareRobustCommand{\VAN}[3]{##3}\VANthebibliography}

\usepackage{graphicx}	
\usepackage{amsmath}	
\usepackage{ulem}
\usepackage{xcolor}

\usepackage{multicol}
\usepackage{graphicx}
\usepackage{subcaption}

\usepackage{natbib}

\newcommand{\hessj}{HESS\,J0632+057}

\title[Orbital parameters of HESS J0632+057]{New insight into the orbital parameters of the gamma-ray binary HESS\,J0632+057}

\author[N. Matchett and B. van Soelen]{
Natalie Matchett$^{1}$\thanks{E-mail: matchettn@ufs.ac.za, vansoelenb@ufs.ac.za}
and Brian van Soelen$^{1}$
\\
$^{1}$Department of Physics, University of the Free State, 205 Nelson Mandela Dr., Bloemfontein, 9300, South Africa
}

\date{Accepted XXX. Received YYY; in original form ZZZ}

\pubyear{2024}

\begin{document}
\label{firstpage}
\pagerange{\pageref{firstpage}--\pageref{lastpage}}
\maketitle

\begin{abstract}
The gamma-ray binary HESS\;J0632+057 consists of a Be star and an undetected compact object in a $\sim$317 day orbit. The interpretation of the emission from this system is complicated by the lack of a clear orbital solution, as two different and incompatible orbital solutions were obtained by previous radial velocity studies of this source.
In order to address this, we report on 24 new observations, covering $\sim$60 per cent of the orbit which we have undertaken with the Southern African Large Telescope (SALT). We obtained new radial velocity measurements from cross-correlation of the narrower spectral features, and by fitting Voigt profiles to the wings of the Balmer emission lines. 
Additionally, we find an indication of orbital variability in the equivalent widths and V/R of the Balmer lines. Using the combined data from this study, as well as archival data from the earlier radial velocity studies, we have derived updated orbital solutions. Using reported H\,$\alpha$ emission radial velocities - previously not considered for the orbital solution - along with the new SALT data, a solution is obtained where the brighter peak in the X-ray and gamma-ray light curves is closer to periastron.
However, continuing sparse coverage in the data around the expected phases of periastron indicates that the orbital solution could be improved with further observation.
\end{abstract}

\begin{keywords}
binaries: spectroscopic -- stars: emission-line, Be -- stars: individual: HESS J0632+057 -- gamma-rays: stars.
\end{keywords}



\section{Introduction}
Gamma-ray binaries are a rare subclass of high-mass binaries, where the majority of their non-thermal emission is produced at energies $>$1 MeV in a $\nu F_{\nu}$ distribution \citep[see e.g.][]{Dubus2013,Masha2019Overview}. In all nine of the known gamma-ray binary systems, the companion star has been identified as either an O type star -- LS\;5039 \citep{Clark2001}, 1FGL\;J1018.6-5856 \citep{FermiList2_2012}, LMC\;P3 \citep{Corbet2016}, 4FGL\;J1405.1-611 \citep{Corbet2019} and HESS\;J1832-093 \citep{vanSoelen2024} -- or a Be-type star -- PSR\;J2032+4127 \citep{Massey1991}, PSR\;B1259-63 \citep{Johnston1994}, LS\;I\;+61$^{\circ}$303 \citep{Gregory1978} and HESS\;J0632+057 \citep{Hog2000} -- with a compact object in the mass range of a neutron star or black hole.
Two long-standing scenarios, involving either a pulsar-wind or microquasar, are routinely proposed to explain the very high energy (VHE) emission observed from these sources \citep{Dubus2013}.
In the pulsar-wind scenario, the compact object is a young neutron star producing a relativistic pulsar wind, which collides with the stellar wind from the O/Be companion.
This forms a stand-off shock, wherein particles are accelerated to very high energies. The non-thermal emission is then produced as the particles cool-down via synchrotron and inverse Compton processes in the shock.
The microquasar scenario suggests that the VHE emission originates from particles accelerated to very high energies in a relativistic jet, which is produced via accretion on to the compact object.
To date, the compact object has been identified as a young pulsar in three systems, namely PSR\,B1259-63 \citep{Johnston1992}, PSR\,J2032+4127 \citep{Lyne2015}, and LS\,I\,+61$^\circ$303 \citep{Weng2022}. The possible detection of pulsed emission has also been reported for LS~5039 \citep{Yoneda2020,Makishima2023} but has not been confirmed \citep{Kargaltsev2023}.

The gamma-ray binary HESS J0632+057, situated near the Rosette Nebula, consists of a B0Vpe star (MWC\,148) and an undetected compact object in a $ P_{\rm orb} = 317.3 \pm 0.7$\,d orbit \citep[derived from X-ray observations;][]{Adams2021}.
Over the course of long-term X-ray and $\gamma$-ray monitoring, the X-ray and TeV light curves consistently displayed two maxima: the primary, sharper peak, occurs around phases $\phi \sim 0.3$--$0.4$ \citep[time of phase zero $T_0 = 2454857.5~\rm{HJD}$;][]{Falcone2010}, while the secondary, flatter peak, occurs around phases $\phi \sim 0.6$--$0.8$ \citep[e.g.][]{Adams2021}.
The interpretation of the multiwavelength emission from \hessj\ is complicated by the two different and incompatible orbital solutions reported by \cite[][hereafter referred as C12]{Casares2012} and \cite[][hereafter referred as M18]{Moritani2018}.
The C12 solution was derived from radial velocity (RV) measurements of the broad, weak absorption lines, phase-folded on the initial orbital period estimate of $P \sim321$\,d \citep{Bongiorno2011}.
This orbital solution would place the phases coinciding with the two maxima in the light curves around apastron, which would be difficult to interpret.
In the later study by M18, the RV measurements were instead obtained from the H~$\alpha$ emission line, and phase-folded on orbital periods of 308 and 313 d, derived from an independent periodicity search by the authors of  optical and X-ray data. The M18 orbital solution would result in the primary X-ray/TeV peak occurring after apastron, while the secondary flatter peak occurs near periastron. They suggested that the maxima in the light curve are a result of the pulsar interacting with the denser equatorial circumstellar disc of the Be star, which is inclined with respect to the orbital plane, similar to what is proposed for PSR B1259-63 \citep[see e.g.][]{Chernyakova2015,Chernyakova2021_ClumpyDisc}.
However, given that the optical emission lines arise from the circumstellar disc and not directly from the photosphere of the B star, the line profile is affected by the structure and behaviour of the disc. Thus, any asymmetry or inhomogeneities in the disc could affect the RV measurements. 
In order to try to mitigate this, M18 measured their velocities from the H~$\alpha$ emission wings using the bisector method \citep{Shafter1986}. Since the wings arise from the disc's inner region, nearest to the Be star surface, it is suggested that this region would be less affected by interactions with the pulsar, and would more closely follow the motion of the Be star.

Given the very different orbital solutions found, the interpretation of the observed multiwavelength emission has been discussed with reference to both the C12 and M18 solutions.
Because of this uncertainty, in addition to the C12 and M18 orbital solutions, several other attempts to constrain the orbital parameters through emission modelling have been undertaken \cite[see][]{Malyshev2019,Tokayer2021,Chen2022,Kim2022}. These orbital solutions are primarily based on assumptions concerning the origin of the observed X-ray/TeV flux peaks, either assuming the peaks are due to the pulsar-disc crossings \citep{Malyshev2019,Chen2022} or that the primary and secondary peaks are as a result of the intra-binary shock (IBS) peaking at periastron, and Doppler beaming along the cometary tail, respectively \citep{Tokayer2021,Kim2022}.

Given the uncertainty in establishing the basic geometry of this binary, here we consider the possible orbital solutions, combining C12 and M18's earlier results with new observations covering 60 per cent of the orbit of \hessj/MWC~148, obtained with the Southern African Large Telescope (SALT).
We discuss how this affects the interpretation of the orbital solutions, as well as the possible implications for the non-thermal emission.

This paper is structured as follows. In Section~\ref{sec:observations}, we present the optical spectroscopic observations obtained with SALT, the different methods implemented to measure the RVs, as well as the equivalent widths (EWs) and V/R ratios of the Balmer emission lines; in Section \ref{sec:Discussion}, we present the results from the SALT data and the implication of the new SALT data on the C12 and M18 orbital solutions, the behaviour of the circumstellar disc and the multiwavelength emission. Finally, the conclusions are presented in Section \ref{sec:Conclusions}.

\section{Observations}
\label{sec:observations}
\subsection{SALT observations and data reduction}
\label{sec:Obs}

Optical spectroscopic observations were undertaken using the High Resolution Spectrograph \citep[HRS;][]{Barnes2008} on SALT \citep{Buckley2006} in HR mode ($R \sim 65\,000$).
The fibre-fed echelle spectrograph separates both the \textit{sky} and \textit{object} fibres into a blue-arm (3700--5550\,\AA) and red-arm (5550--8900\,\AA) which then undergo cross-dispersion on to the blue/red cameras.

\begin{figure*}
\includegraphics[width=\textwidth]{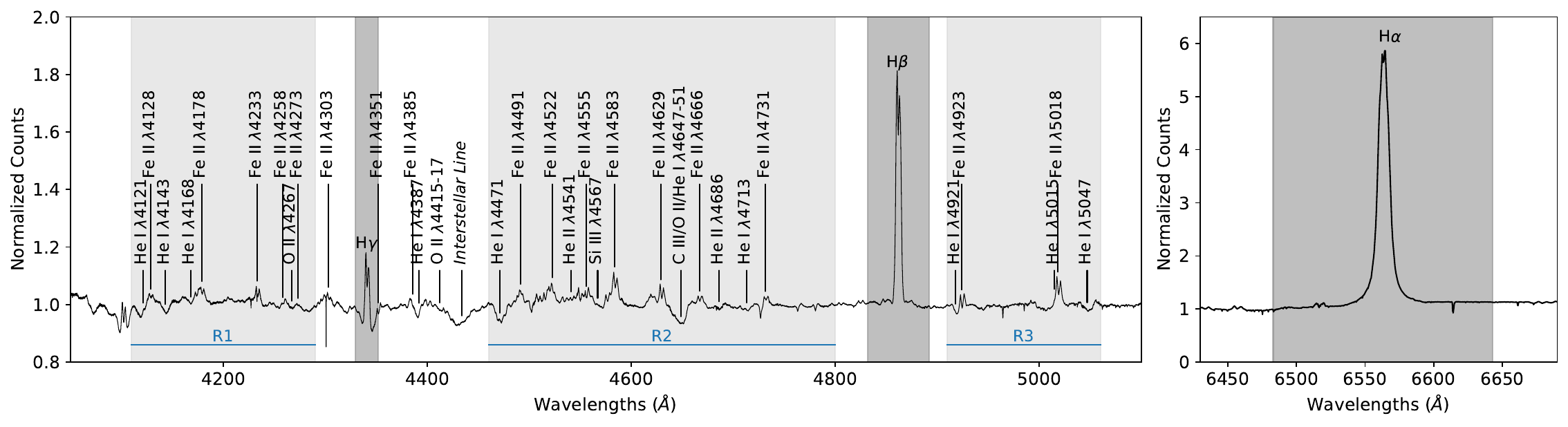}
\caption{A representative blue arm template spectrum (left) showing the H~$\beta$ and H~$\gamma$ Balmer emission lines as well as the low intensity Fe\,{\sc ii} lines and weak absorption features for reference, and a portion of the averaged red arm spectrum (right) showing the H~$\alpha$ emission line. The dark grey shaded regions on the blue and red spectra indicate the regions around the Balmer emission lines included in the Voight profile fits (Section \ref{sec:VoigtFitting}). The light grey regions in the blue spectrum show the three regions (R1, R2 and R3) that were used for cross-correlation (Section \ref{sec:CC}).}
\label{fig:Spec}
\end{figure*}

Twenty-four observations consisting of $3 \times 600$\,s (first semester) or $3\times 700$\,s (subsequent semesters) exposures were obtained over the course of three observing semesters, between 2020 December 7 and 2023 March 11. The individual echelle orders of the \textit{sky} and \textit{object} fibres were extracted and wavelength calibrated by the SALT-HRS pipeline \citep{Kniazev2016}.
The individual orders of the echelle spectra were then merged, with special care taken to minimize the noise where the different orders overlapped, as many of the broader spectral features extended across more than one echelle order.
The merged \textit{sky} fibre spectra were then subtracted from the corresponding merged \textit{object} fibre spectra, before being normalized, and rebinned to ensure a uniform wavelength dispersion across the entire spectrum. Finally, the spectra were corrected to the barycentre using the standard procedures in \textsc{iraf/pyraf}, and a single nightly averaged spectrum was produced from the three exposures obtained each night, for both arms.
Fig.~\ref{fig:Spec} shows portions of the obtained blue- and red arm spectra, averaged over all observations. The spectrum of MWC\,148 consists of strong Balmer emission lines -- with either a fine structure at the line peak (H~$\alpha$) or a double peak structure (H~$\beta$ $\&$ H~$\gamma$) -- and several low intensity Fe\,{\sc ii} emission lines arising from the circumstellar disc of the Be star, in addition to the broad, weak absorption lines produced in the photosphere of the star.

\subsection{Radial velocity measurements}
\label{sec:RV_Measurements}

In general, RV measurements from a star are ideally obtained from the photospheric absorption lines. However, as will be discussed in Section~\ref{sec:CC}, the contribution from the circumstellar disc complicates these measurements for Be stars.

In this work, the RV measurements were obtained through two different methods. First, by fitting Voigt profiles to the wings of the H~$\alpha$, H~$\beta$ and H~$\gamma$ Balmer emission lines, and, second, through cross-correlation of regions of the spectra excluding the strong Balmer emission lines. Obtaining RV measurements from several different line species, corresponding to different regions across the circumstellar disc, should help mitigate any local variation within these different emission regions.

\subsubsection{Radial velocity from the Balmer emission lines}
\label{sec:VoigtFitting}

We have made use of Voigt profile fitting to the wings of the H~$\alpha$, H~$\beta$ and H~$\gamma$ Balmer emission lines to obtain the RV measurements, similar to the method in the earlier study by \cite{Moritani2015}.
The wings are produced by the more central regions of the circumstellar disc, which should be less affected by any variability present in the disc from tidal interaction with the companion.
We made use of the \texttt{Voigt1D} model in the \texttt{astropy} package \citep{astropy:2013}, in order to obtain a best-fitting to the emission-line data obtained with the standard least-squares procedure in \texttt{scipy}.
 
For each nightly averaged spectrum, the procedure for the Voigt fitting was as follows: 

(i) A region around each emission line was chosen for the Voigt fitting which excluded other spectral features, but still included sufficient continuum (dark shaded regions in Fig.~\ref{fig:Spec}).

(ii) Inspection of the spectra surrounding the emission lines showed a slight slope in the continuum around the H~$\alpha$ and H~$\beta$ lines, while the H~$\gamma$ emission-line emerges from the underlying absorption feature. Therefore, an additional continuum correction was performed on the selected wavelength region.
For the H~$\alpha$ and H~$\beta$ emission lines, a first order polynomial gives a good fit to the continuum.
In the case of the H~$\gamma$ emission line, a fourth order polynomial was used to fit the continuum and the underlying absorption feature from the star.

(iii) Finally, the Voigt fitting was performed in a two-step process: First, an initial fit to each emission profile was performed, which excluded the central $\sim$5 \AA\ region of the line, to avoid the double peak profile. Secondly, the Voigt profile was re-fit to only the wings of the emission line, which was defined as the data falling below 25 per cent of the peak of the initial fit.

\begin{figure}
\includegraphics[width=\columnwidth]{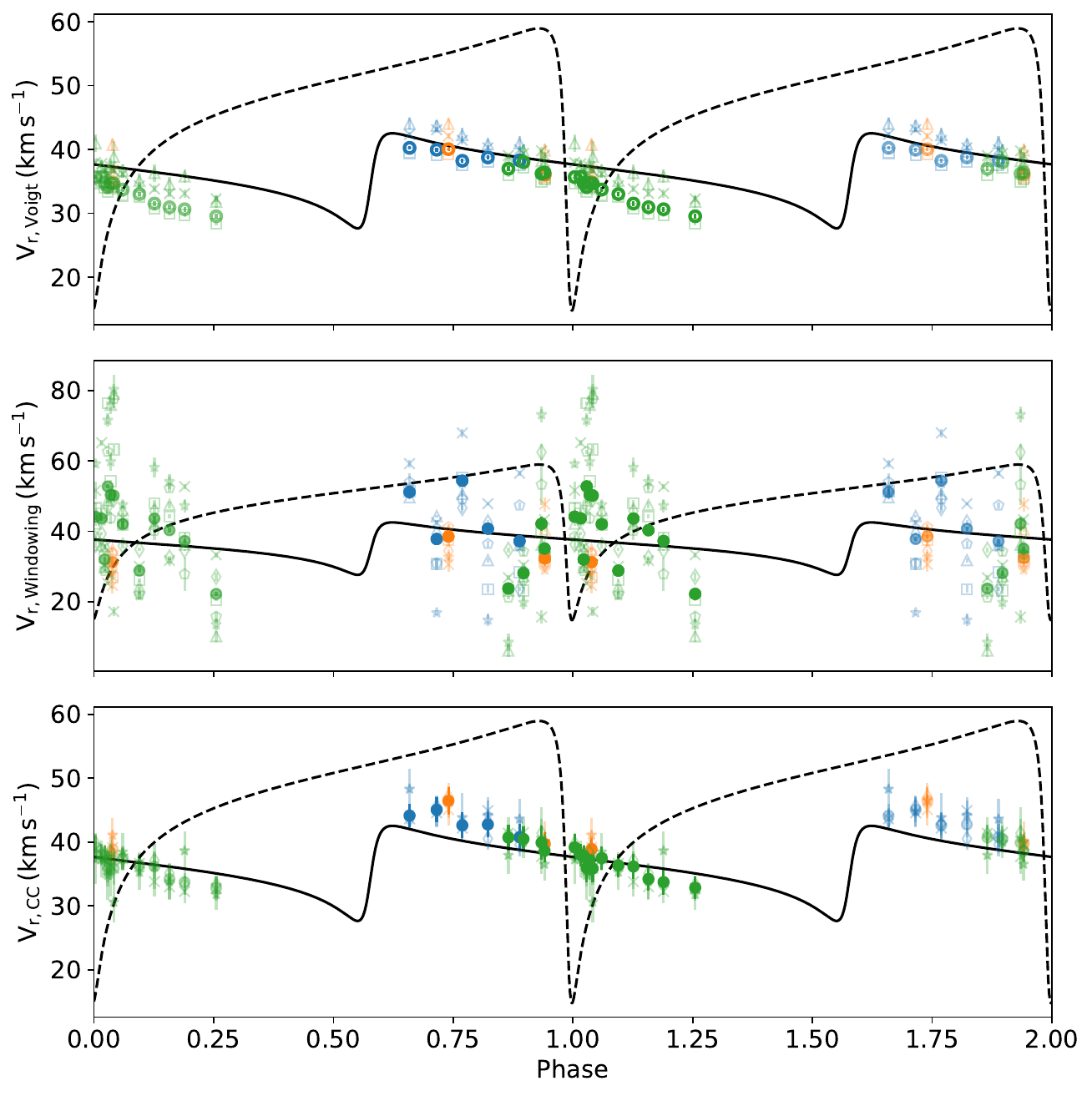}
\caption{SALT RVs versus phase plotted over the existing C12 (dashed line) and M18 (solid line) orbital solutions, refolded on $P_{\rm orb} = 317.3$\,d. The results are repeated over two orbital phases for clarity and observations during the first, second and third semester are marked as blue, orange and green, respectively. The error bars indicate the statistical error.
{\it Top:} The RV measurements from the H~$\alpha$ (crosses), H~$\beta$ (open squares) and H~$\gamma$ (open triangles) emission lines, and weighted average (open circles). 
{\it Middle:} The RVs as found via cross-correlation of spectral windows around the He\,{\sc i} $\lambda$4121 (crosses), $\lambda$4387 (triangles), $\lambda$4471 (squares), $\lambda$5047 (diamonds), Si\,{\sc iii} $\lambda$4567 (stars), and C\,{\sc iii}/O\,{\sc ii}/He\,{\sc i} blend (4647--4651\,\AA{}; pentagons) absorption features. The weighted average of the RVs from all the absorption lines are indicated by the solid circles.
{\it Bottom:} The RV measurements for the different cross-correlation regions R1 (crosses), R2 (open triangles) and R3 (open squares), and weighted average (filled circles).
}
\label{fig:RVs}
\end{figure}

The RVs were obtained from the centre of the Voigt profile for each of the three Balmer emission lines, from which the weighted averages for the RV were calculated. The RV for each line as well as the average are shown in the top panel of Fig.~\ref{fig:RVs}. The average radial velocities, used in all further calculations, are listed in Table \ref{table:RVs} and are shown (red, open circles) folded on $P_{\rm{orb}} = 317.3$\,d in the top panel of Fig.~\ref{fig:SALT+solns}.

\subsubsection{Radial velocity from cross-correlation}
\label{sec:CC}

For MWC 148, cross-correlation against the photospheric absorption lines is complicated by the Balmer emission lines, which originate from the circumstellar disc, and fill-in the absorption lines. Additionally, there are several other, lower intensity emission lines in the spectra, e.g.\ Fe~{\sc ii}. The remaining absorption lines are weaker and very broad due to rotational broadening.

We first attempted to measure the RVs from the broad absorption lines, by performing cross-correlation on spectral windows covering the individual He\,{\sc i} ($\lambda$4121, $\lambda$4387, $\lambda$4471, $\lambda$5047), Si\,{\sc iii} ($\lambda$4567), and C\,{\sc iii}/O\,{\sc ii}/He\,{\sc i} blend  (4647--4651\,\AA{}) features. Crucially, this is similar to the method described in C12. The cross-correlation was performed using the \textsc{xcsao} package \citep{Kurtz1998} in \textsc{iraf/pyraf}, against a template spectrum produced from the average of the available blue-arm observations, similar to the process used by e.g.\ \cite{Foellmi2003}, \cite{Monageng2017} and \cite{vanSoelen2019}.
This allowed us to directly correlate the spectra with a template that matches the target and which is taken with the same instrumental configuration. However, this also introduces any instrumental effects into the template (e.g. the noise at the merger sites of the orders in echelle spectra) and causes the spectral features to be artificially broadened due to the RV off-set of each frame. 
In order to address the broadening, the template was created in a two step process: first the RV off-set between each spectrum and the average of all spectra was determined via cross-correlation; and second, the template was then produced by averaging over all spectra after the initial RV offset had been removed.
The velocity of the template was determined from the weighted average of the RVs of the H~$\beta$ and H~$\gamma$ emission lines, following the method discussed in Section \ref{sec:VoigtFitting}. The RVs obtained for each spectral window via cross-correlation, as well as the average, are shown in the middle panel of Figure \ref{fig:RVs}. Note, the errors shown are only the statistical errors obtained from the cross-correlation analysis. The RV curve obtained by C12 (re-folded on $P_{\rm orb} = 317.3$\,d) is shown by the dashed line. The new observations do not show an indication of the sharp minimum associated with their fit.

Given that the attempts to cross-correlate predominantly against the broad absorption lines produced poor results, we focused on the narrower, lower intensity features in the spectra which are formed deeper in the photosphere and are less affected by the strong stellar wind, minimizing some of the scatter. These spectral regions also included narrow Fe lines which, while arising from the circumstellar disc, originate closer in than the Balmer lines \citep{Arias2006,Zamanov2015,Zamanov2016,Moritani2015}.
From the average peak separation of the emission lines in the SALT spectra we estimate the radii of the H~$\alpha$, H~$\beta$, H~$\gamma$ and Fe~{\sc ii} emission regions in the disc to be $\sim$ $52R_*$, $\sim17R_*$, $\sim15R_*$, and $\sim$10--13$R_*$, respectively, where the stellar radius was taken as $R_*=6.6\,{\rm R}_\odot$. The orbital separation at periastron, based on the existing C12 and M18 orbital solutions, ranges between 13.2 and 18\,$R_*$. Thus, at least, the Fe {\sc ii} emission region, up to the H~$\beta$ emission region, is within the separation distance and would be less affected by any tidal interaction with the compact object.

The cross-correlation was performed separately on three regions of the blue spectrum, labelled R1 (4110--4290 \AA), R2 (4460--4800 \AA), and R3 (4910--5060 \AA) in Fig.~\ref{fig:Spec}. These regions were chosen to exclude the Balmer emission lines and the broad interstellar line at $\sim4425$~\AA. A template was produced following the process detailed above for each of the three cross-correlation regions. The {\sc xcsao} package first determines the Fourier transform of the object spectrum and the conjugate of the Fourier transform of the template spectrum, before finding the cross-correlation from the product between the two \citep{Kurtz1998}. Before this is performed, a section of the high and low frequencies in the Fourier space are removed. This is typically undertaken so as to filter out noise (high frequency) and broader continuum features (low frequency) that are still present in the spectra \citep{Kurtz1998}. In our final cross-correlation, we increased the level of the low-frequency filtering. This had the effect of minimizing the relative strength of the broad, weak absorption lines, while increasing the relative strength of the narrow features in the cross-correlation. This resulted in far less scatter in the RV measurements.

The RV measurements obtained from the cross-correlation for each of the three regions, as well as the weighted average, are shown in the bottom panel of Fig.\;\ref{fig:RVs}. The weighted average of the RVs, used for all further discussions, are given in Table~\ref{table:RVs}, and are shown (blue, filled circles) folded on $P_{\rm orb} = 317.3$\,d in the top panel of Fig.~\ref{fig:SALT+solns}.

\begin{table}
\centering
\caption{The average RV measurements obtained through the Voigt profile fitting to the Balmer emission lines, and via cross-correlation of regions R1, R2 and R3}.
\begin{tabular}{c c c}
\hline
Observation Date & $V_{\rm{r},\left<Voigt\right>}$ & $V_{\rm{r},\left<CC\right>}$ \\
BJD & km\,s$^{-1}$ & km\,s$^{-1}$\\ 
\hline
2459191.4631 & 40.2 $\pm$ 0.3 & 43.8 $\pm$ 1.8\\
2459209.4354 & 40.0 $\pm$ 0.3 & 44.6 $\pm$ 2.0\\
2459226.4149 & 38.2 $\pm$ 0.3 & 42.2 $\pm$ 2.1\\
2459243.3574 & 38.7 $\pm$ 0.3 & 42.4 $\pm$ 1.9\\
2459264.3257 & 38.3 $\pm$ 0.4 & 40.3 $\pm$ 2.1\\
2459534.5192 & 40.1 $\pm$ 0.3 & 46.1 $\pm$ 2.1\\
2459598.3477 & 36.0 $\pm$ 0.4 & 39.3 $\pm$ 1.8\\
2459629.3266 & 34.9 $\pm$ 0.3 & 38.4 $\pm$ 1.8\\
2459891.5433 & 37.0 $\pm$ 0.4 & 40.3 $\pm$ 2.0\\
2459901.5435 & 38.0 $\pm$ 0.4 & 40.2 $\pm$ 1.9\\
2459913.4921 & 36.2 $\pm$ 0.5 & 39.6 $\pm$ 2.8\\
2459915.4826 & 36.4 $\pm$ 0.4 & 38.1 $\pm$ 1.8\\
2459935.4242 & 35.7 $\pm$ 0.4 & 38.8 $\pm$ 2.1\\
2459939.4187 & 35.7 $\pm$ 0.4 & 37.7 $\pm$ 1.8\\
2459941.4088 & 34.7 $\pm$ 0.4 & 37.4 $\pm$ 1.6\\
2459943.4068 & 34.1 $\pm$ 0.4 & 35.5 $\pm$ 2.3\\
2459945.4004 & 34.7 $\pm$ 0.4 & 36.7 $\pm$ 2.4\\
2459947.4081 & 34.7 $\pm$ 0.4 & 35.4 $\pm$ 2.2\\
2459953.3960 & 33.7 $\pm$ 0.4 & 37.1 $\pm$ 1.8\\
2459964.4012 & 33.0 $\pm$ 0.4 & 35.8 $\pm$ 1.9\\
2459974.3705 & 31.5 $\pm$ 0.3 & 35.8 $\pm$ 2.0\\
2459984.3359 & 31.0 $\pm$ 0.4 & 33.9 $\pm$ 1.9\\
2459994.3370 & 30.6 $\pm$ 0.4 & 33.5 $\pm$ 2.0\\
2460015.2748 & 29.5 $\pm$ 0.4 & 32.4 $\pm$ 1.8\\
\hline
\end{tabular}
\label{table:RVs}
\end{table}

\subsection{Variation of the emission lines}
\label{sec:em_behaviour}

In addition to the RV measurements, we measured the variability in the EW and the V/R profile of the Balmer emission lines.
The EW measurements were performed on H~$\alpha$, H~$\beta$, and H~$\gamma$, by summing the area under the lines. The error in the EW was calculated following eq.\ 7 in \cite{Vollmann2006}, using a signal-to-noise ratio that was obtained using the method presented by \cite{Stoehr2008}. The H~$\alpha$ and H~$\beta$ EW values, when phase-folded, show evidence of an orbital modulation. There is also an indication of orbit-to-orbit variation, with the first semester observations being slightly larger than the second (see around phase 0.8). The trend in the phase-folded data shows a lower $|{\rm EW}|$ between $\phi \sim 0-0.25$, before a gap in the data not covered by the SALT observations, around the predicted phase of the periastron. The $|{\rm EW}|$ strength at $\phi \sim 0.6$ is higher and steadily declines towards the `low' state again.

\begin{figure}
\includegraphics[width=\columnwidth]{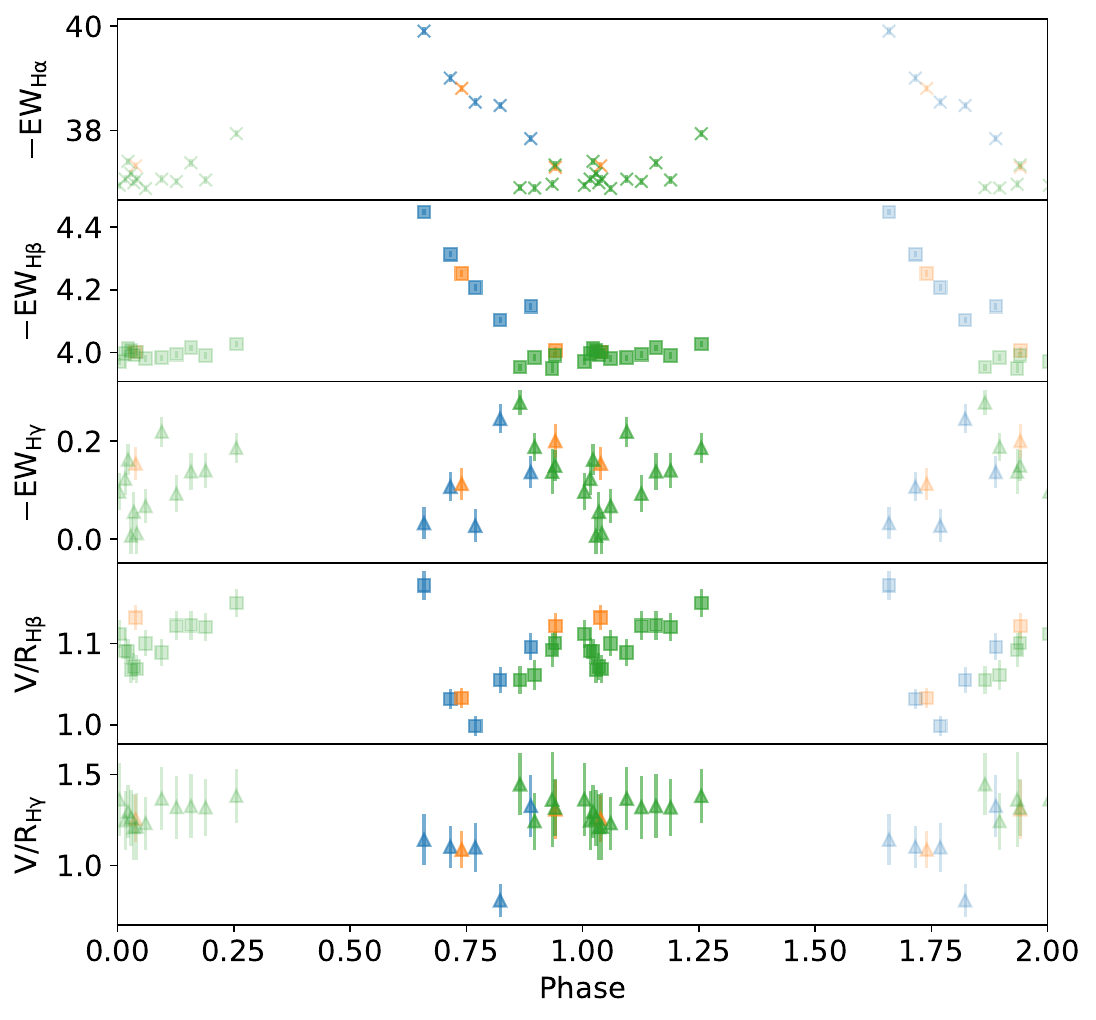}
\caption{From top to bottom: the EW measurements for the H~$\alpha$, H~$\beta$ and H~$\gamma$ emission lines, along with the V/R ratio for the H~$\beta$ and H~$\gamma$ emission lines, folded on the 317.3~d period. The observations from the first, second, and third observing semesters are shown in blue, orange, and green, respectively. The data are repeated over two orbits for clarity.}
\label{fig:EW+VR_Phase}
\end{figure}

The H~$\beta$ and H~$\gamma$ lines show a typical double-peaked structure associated with emission from a disc. In order to measure the variation in the relative strength of the blue-to-red peak (known as a V/R profile) we fit the combination of two Voigt profiles, one to each peak, to the entirety of the emission line profiles. 
The continuum correction was performed in the same way as discussed in section~\ref{sec:VoigtFitting}.
The results are shown in Fig.~\ref{fig:EW+VR_Phase}. There is similarly an indication of orbital modulation in the V/R profile of the H~$\beta$ and (to a lesser extent) the H~$\gamma$ line (lower two panels on Fig.~\ref{fig:EW+VR_Phase}). The V/R profile of H~$\beta$ increases from $\approx 1$ to $\approx 1.1$ from phase $\sim 0.7$ to $\sim 1.25$. We also note a large change from the first observation to the second during the first observing semester, indicating a change in the disc symmetry.

\section{Discussion}
\label{sec:Discussion}

\subsection{Comparison between the velocity results}

For our analysis, cross-correlating against the broad, weak absorption lines in the spectrum of the Be companion produces a large scatter in the RVs. Be stars are known to be non-radially pulsating stars \cite[see e.g.][]{Rivinius2003},  which has been suggested to be linked to the mass-loss mechanism, lifting material from the stellar surface to form the circumstellar disc \citep[see][and references therein]{Baade2016, Rivinius2016}. However, this means that the photospheric absorption lines are prone to rapid line profile variations \citep{Rivinius1998b,Rivinius2003}. 
In addition, the atmospheric layers wherein the optical photospheric (absorption) lines are formed in OB stars have outflow velocities of 20-30 $\rm{km \, s^{-1}}$ and are variable due to non-radial pulsations and wind fluctuations \citep{Oijen1989}. The rapid line profile variation of the absorption lines (combined with the low S/N) and the fast, variable stellar outflow adds stochastic variations which significantly reduce their value for dynamical studies in Be stars \citep[see e.g.][]{Prisegen2020}.

In contrast, the RV measurements obtained through the Voigt profile fitting and final cross-correlation produces significantly less scatter with consistent results across the different line species, and over the different observing semesters (orbits). Although interaction between the compact object and the disc may produce tidal disruptions and asymmetric density structures in the disc, the emission line profiles are produced over large regions in the disc and, therefore, small variations are likely to be averaged out across the relevant emitting regions. In addition, in our analysis we relied on several different line species, formed in different regions of the disc, lessening the impact of larger variations that may be present in some of the lines.

The RVs obtained from the SALT spectra are similar to the H~$\alpha$ RV data from M18, as well as those reported in fig. 4 of C12, despite the changes in the H~$\alpha$ line strength.

\subsection{Orbital parameters}

\subsubsection{Comparison with existing solutions}

The average RV from our Voigt fitting (open red circles) and cross-correlation of regions R1, R2, and R3 (filled blue circles), are shown in the top panel of Fig.\ \ref{fig:SALT+solns}. The results are compared to the RV solutions obtained for the C12 and M18 data refolded on the newer orbital period of 317.3\,d. The change in the orbital period does not introduce a significant change to the existing orbital solutions, as the orbital parameters are all within the uncertainties of the previously published solutions (see Table \ref{tab:orbital_params}). However, we note that the eccentricity of the refolded M18 data increases to $e\approx0.8$.

The new SALT RV measurements obtained via these two methods show consistent results, though there is a small offset of $\approx 3$\,km\,s$^{-1}$.
This is likely due to the difference found in the RV of the template used for the cross-correlation.
These RV measurements are in reasonable agreement with the re-folded M18 solution, but are significantly different to the solution found in C12.

\subsubsection{Orbital solution with SALT and M18 data}
\label{sec:M18+SALTsoln}

Since the current SALT observations do not cover the full orbit, we obtained a new orbital solution by combining the M18 RVs and the average of the SALT RVs (i.e. the average found from the Voigt profile fitting and the cross-correlation). The best RV fit is shown by the dotted line in the middle panel of Fig.~\ref{fig:SALT+solns}, while the parameters are listed in Table \ref{tab:orbital_params}, and the orbit is shown in Fig.\,\ref{fig:Orbit}(a).

While the new solution is similar to that found from the M18 data alone, it is important to note that with the new orbital period, there is sparse coverage and a larger scatter between phases $\sim 0.3$--$0.6$, which is not covered by the new SALT observations. Periastron lies between these phases. This means that the key parameters of the phase of periastron, the longitude of periastron, and the eccentricity of the orbit are still very poorly constrained (only the statistical errors from the fit are given in Table~\ref{tab:orbital_params}). Increased phase coverage over these critical phases, could significantly improve the orbital solution.

\subsubsection{Orbital solution with SALT and C12 H~$\alpha$ data}
\label{sec:C12Ha+SALTsoln}

\begin{figure}
\includegraphics[width=0.49\textwidth]{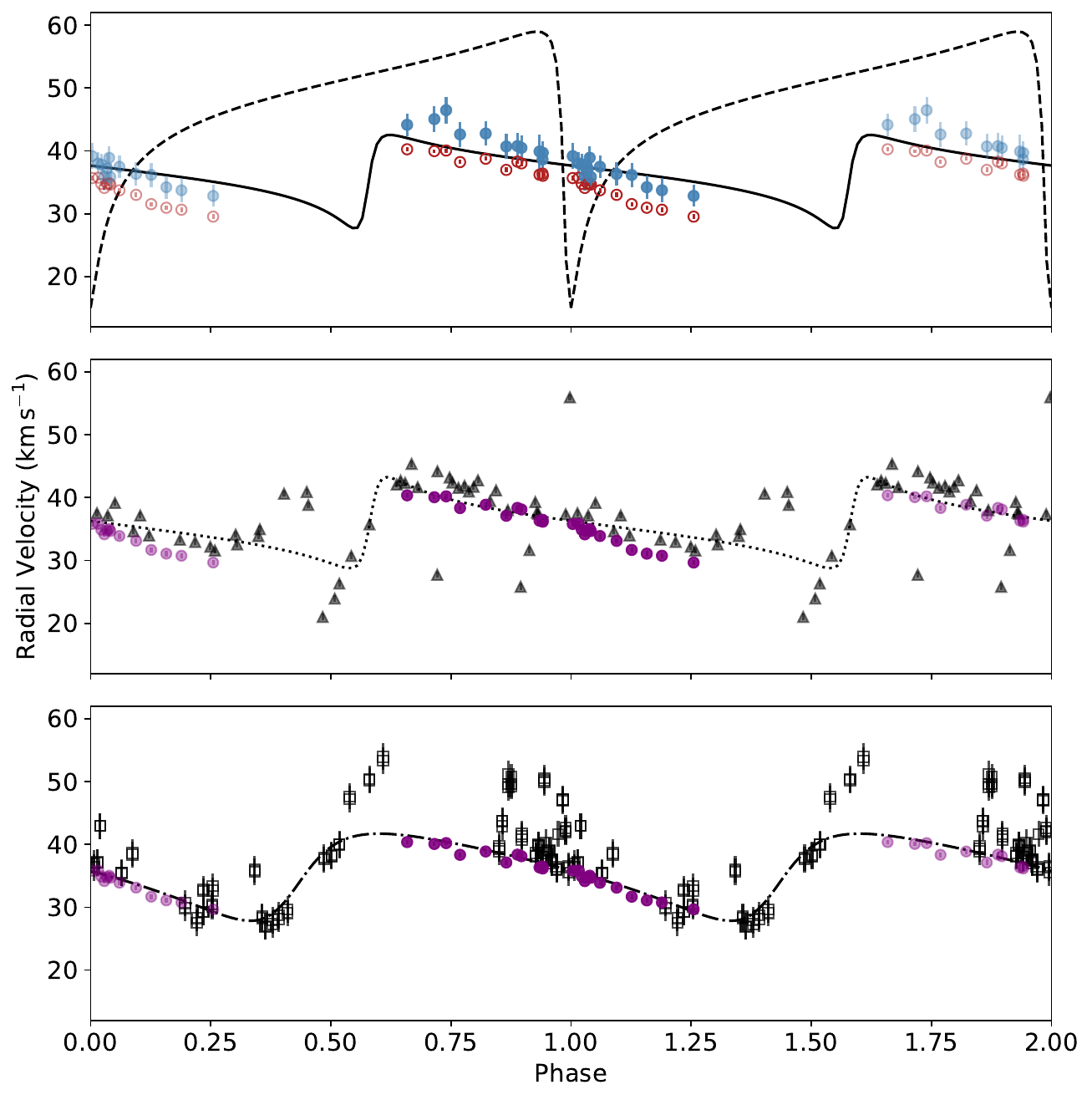}
\caption{Weighted average (top) of our Voigt profile-fitting (red open circles) and cross-correlation (blue filled circles) RV results, plotted against the C12 (dashed line) and M18 (solid line) orbital solutions, refolded on the 317.3\,d period. Our average SALT RV values (purple circles) are fitted with the M18 data (middle) and C12 H~$\alpha$ data (bottom) to provide new orbital solutions (see Sections \ref{sec:M18+SALTsoln} and \ref{sec:C12Ha+SALTsoln} for details). Error bars may be smaller than the data points.}
\label{fig:SALT+solns}
\end{figure}

\begin{figure*}
\begin{subfigure}{0.257\linewidth}
\centering
\includegraphics[width=\linewidth]{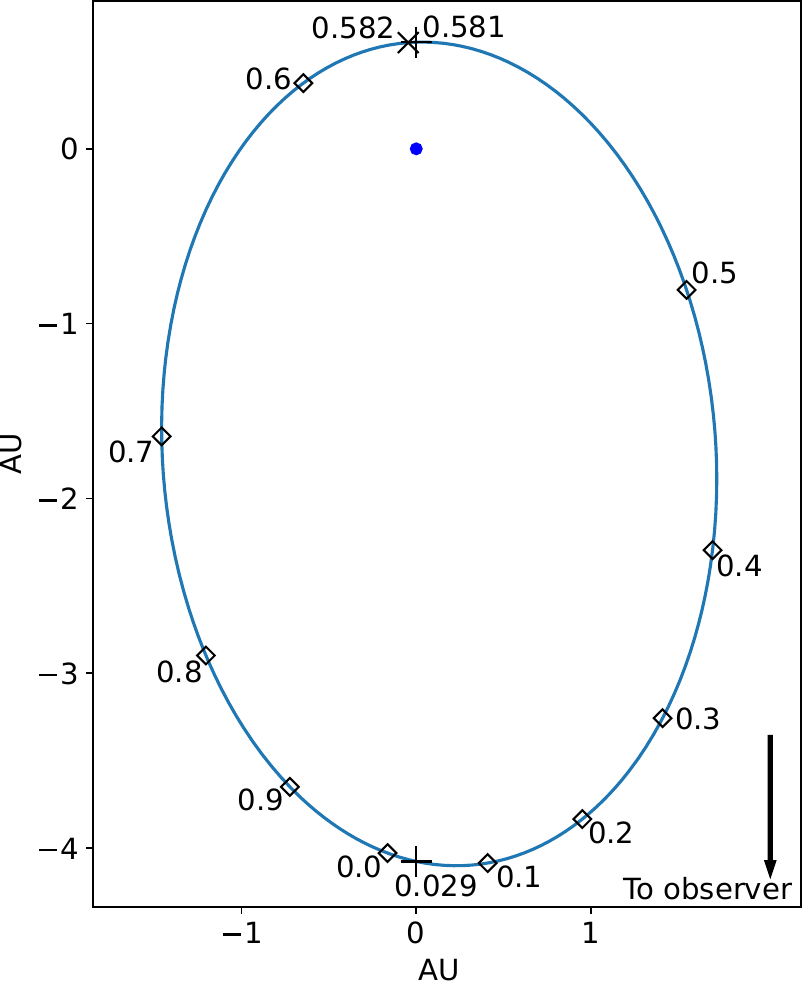}
\caption{}
\label{fig:M18+SALT_orbit}
\end{subfigure}
\qquad \qquad
\begin{subfigure}{0.32\linewidth}
\centering
\includegraphics[width=\linewidth]{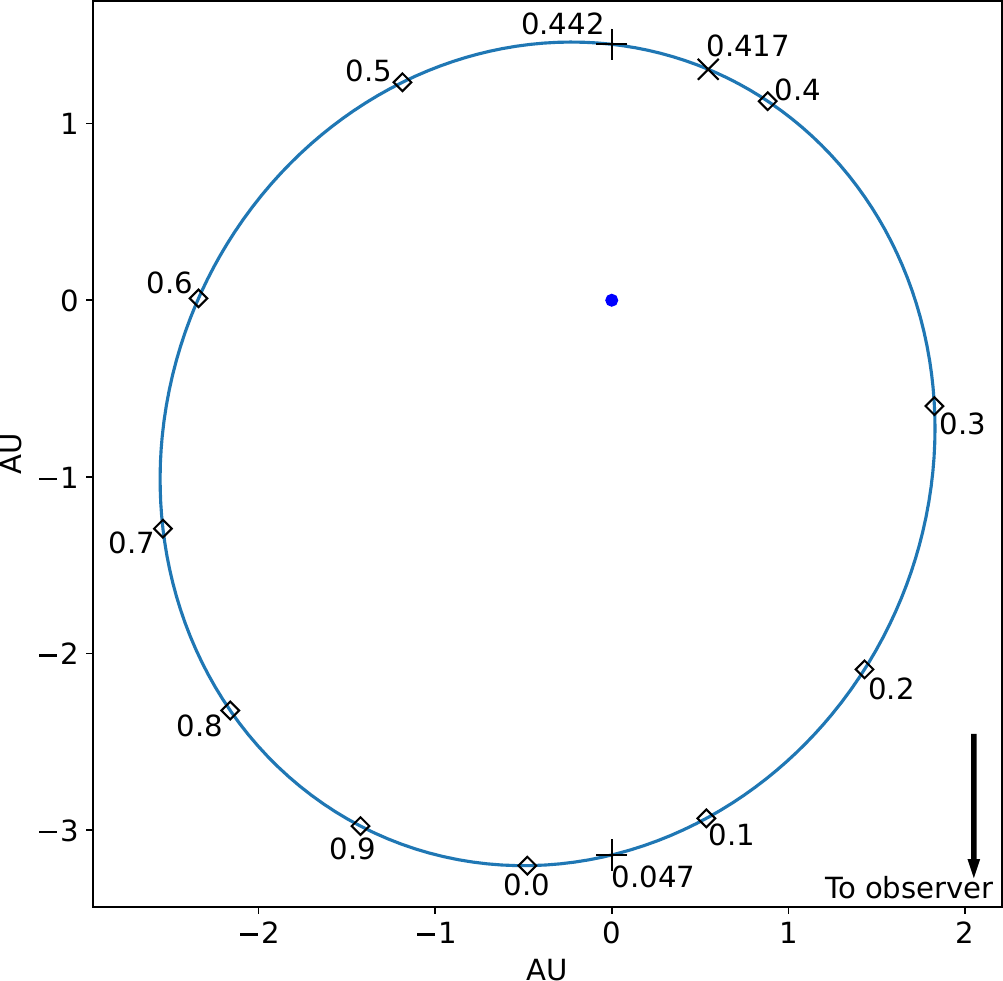}
\caption{}
\label{fig:C12+SALT_orbit}
\end{subfigure}
\caption{The orbital geometry for the best fit to the (a) SALT and M18 RV data and, (b) SALT and C12 H~$\alpha$ data -- assuming masses of 16\,M$_{\odot}$ and 1.4\,$M_{\odot}$ for the Be star \citep{Aragona2010} and the compact object, respectively. The Be star (MWC~148) is indicated by the blue circle. Superior and inferior conjunction are indicated with a $+$ and periastron is indicated with a $\times$. The position of the observer is indicated by the arrow.}
\label{fig:Orbit}
\end{figure*}

Given the sparse coverage and large scatter in the M18 data around the expected phases of periastron, we have also extracted the RVs reported in fig.\ 4 of C12, which were obtained from their Liverpool Telescope (LT) spectra of H~$\alpha$. However, we note the large offset between the LT RVs and the spectra obtained with the William Herschel Telescope, Mercator Telescope and STELLA-I Telescope also shown in fig.\ 4 of C12, as well as with respect to SALT. This is likely due to the re-binning that was performed on the LT spectra by C12. 
In order to remove the systematic offset, we have subtracted the average off-set between the C12 and SALT data -- weighted by the scatter of the points in phase bins of width 0.1. In addition, given that the uncertainties in the RV measurements are smaller than the data points, we have estimated the uncertainties to be equivalent to the size of the data points. We then obtained a further orbital solution by combining these data with our average SALT RVs as above. The orbital parameters found are given in Table~\ref{tab:orbital_params}, the best RV fit is shown in the bottom panel of Fig.~\ref{fig:SALT+solns}, and the orbit is shown in Fig.~\ref{fig:Orbit}(b).

This orbital solution is much less eccentric ($e=0.4$) and places the phase of periastron at $\phi \approx 0.42$, which is earlier than is obtained with the M18 data. There is, however, still a large scatter around certain phases, but the RVs follow a more consistent trend compared to the previous results.

\begin{table*}
    \centering
        \caption{Orbital parameters for \hessj\ from our updated RV solutions, combined with C12 and M18 data.$^{\rm a}$}
    \label{tab:orbital_params}
    \begin{tabular}{lcccc} \hline
    Parameter & C12 & M18 & $\rm{SALT_{\left< Voigt+CC \right>}}$ + M18 &  $\rm{SALT_{\left< Voigt+CC \right>}}$ + $\rm{C12_{H\alpha}}$ \\ \hline
      ${P_{\rm{orb}}}^{\rm b}$ (Orbital period; d) & 317.3 & 317.3 & 317.3 & 317.3 \\
      $\rm{T_{peri}}$ (time of periastron; BJD) & 2455488.8 $\pm$ 2.6 & 2454405.4 $\pm$ 5.6 & 2455042.2 $\pm$ 4.0 & 2454989.8 $\pm$ 9.4 \\
      $\phi_{\rm{peri}}$ (phase of periastron) & 0.990 $\pm$ 0.008 & 0.575 $\pm$ 0.018 & 0.582 $\pm$ 0.013 & 0.417 $\pm$ 0.030 \\
      $e$ (eccentricity) & 0.83 $\pm$ 0.08 & 0.76 $\pm$ 0.29 & 0.75 $\pm$ 0.24 & 0.40 $\pm$ 0.08\\
      $\omega$ (longitude of periastron; $^\circ$) & 128.5 $\pm$ 18.8 & 255.5 $\pm$ 26.1 & 274.0 $\pm$ 19.9 & 247.3 $\pm$ 10.1 \\
      $K$ (semi-amplitude; $\rm{km \; s^{-1}}$) & 22.1 $\pm$ 5.8 & 7.5 $\pm$ 4.9 & 7.3 $\pm$ 3.9 & 6.9 $\pm$ 0.8 \\
      $\gamma$ (systemic velocity; $\rm{km \; s^{-1}}$) & 48.3 $\pm$ 1.9 & 36.5 $\pm$ 0.8 & 35.6 $\pm$ 0.6 & 35.8 $\pm$ 0.4 \\ 
      $a \sin i$ (AU) & 0.364 $\pm$ 0.121 & 0.142 $\pm$ 0.118 & 0.142 $\pm$ 0.090 & 0.185 $\pm$ 0.021\\
      $f$ (mass function; M$_{\odot}$) & 0.344 $\pm$ 0.279 & 0.014 $\pm$ 0.027 & 0.013 $\pm$ 0.019 & 0.011 $\pm$ 0.004\\
        \hline
    \end{tabular} \\
\flushleft
\footnotesize{$^{\rm a}$ The uncertainties are the statistical errors found from the fit. In order to account for systematic uncertainties, the reported errors on the parameters are found by scaling the error in the RVs during the fitting, in order to achieve a fit with a reduced $\chi^2$ value of 1 \citep{Lampton1976}. \\
$^{\rm b}$ Fixed at this value.}
\end{table*}

\subsection{Behaviour of the circumstellar disc}

The average EW of the H~$\alpha$ line in this work is $|{\rm EW}| = 37.6$\,\AA. This is stronger than was found by \citet{Moritani2015} ($|{\rm EW}|\approx30$\,\AA), but is less than the maximum seen by \citet{Aragona2010} and C12, thereby confirming its long-term variation. We find an indication of orbital modulation of the EWs, unlike what was  previously found in \cite{Adams2021}, but similar to what was reported by C12. The trend observed in the EW and V/R in this work would be consistent with the variations in the disc induced by tidal interaction between the compact object and the Be star around periastron, introducing an asymmetry to the disc, as indicated by the V/R profile. This is similar to what has been observed for other gamma-ray binaries which contain Be-type stars \citep[e.g.][]{Chernyakova2020_2032,Chernyakova2021_ClumpyDisc}.

\subsection{Implications for the multi-wavelength emission}

The discrepancy between the C12 and M18 solutions has led to difficulty in interpreting the behaviour of \hessj. The new RVs reported here, along with the new orbital solutions determined with the updated orbital period and the H~$\alpha$ RV data from M18 and C12, respectively, strongly suggest periastron lies between orbital phases $\phi \approx 0.3$--$0.6$, in contrast to the C12 solution. The orbital solution from the combined SALT and M18 data would place the first peak in the X-ray/TeV light curves ($\phi\approx 0.3$), which is brighter and narrower,  after apastron and the broader, fainter second X-ray/Tev peak, at $\phi \approx 0.6$--$0.8$, after periastron.

However, the alternative solution, obtained with the SALT +  C12 H\,$\alpha$ data, places the first peak closer to periastron, and the second peak closer to apastron (for a much lower eccentricity). Interestingly, this solution would be in line with that suggested by \citet{Malyshev2019}, though the longitude of periastron is different. From X-ray observations, the authors showed that the hydrogen column density, $N_{\rm{H}}$, produces a larger, primary peak at $\phi\sim0.35$ and a smaller, secondary peak at $\phi\sim0.7$ \citep[see][and references therein]{Malyshev2019}. Since the density of the circumstellar disc decreases with distance from the star, while the scale height increases, the second orbital solution we find would be consistent with this result for an inclined disc, if the emission region is close to the compact object. In this scenario, the peaks in the X-ray/Tev light curves are produced during the pulsar-disc crossing, as has been previously suggested. The variation in the strength of the emission lines indicates orbit-to-orbit variability of the circumstellar disc. A change in the disc density would change the interaction between the compact object and the circumstellar disc and lead to the orbit-to-orbit variability observed within the non-thermal emission \citep[e.g.][]{Adams2021}.

\section{Conclusions}
\label{sec:Conclusions}

We present new, independent RV measurements obtained with SALT for the gamma-ray binary \hessj\ where two different, inconsistent orbital solutions have previously been found. The RVs are calculated from Voigt profiles fit to the Balmer emission-line wings, and from cross-correlation of filtered regions of the spectra which contained narrow absorption features and low-intensity Fe\,{\sc ii} emission lines. A search for RVs from the absorption lines alone did not find a consistent result.

The SALT data are more consistent with the M18 solution. From the SALT data alone, we can constrain the period of periastron to lie between the phases of 0.3-0.6, which disagrees with the phase of periastron obtained in the C12 solution.
We present new orbital solutions from the combination of the new SALT observations and the M18 data, which finds a similar solution to M18, but is poorly constrained around periastron.
However, combining the SALT observations with the RVs reported for the H\,$\alpha$ emission by C12 we find a less eccentric orbital solution which would provide a better interpretation of the X-ray and TeV light curves as the pulsar crossing an inclined disc.

Furthermore, we found orbital modulation in the EW which suggests that there is variation in the circumstellar disc and/or disruption of the disc caused by interaction with the compact object. We suggest variation in the circumstellar disc may be linked to the orbit-to-orbit variability reported in the non-thermal light curves.

Additional observations around phases $0.3$--$0.6$, not covered by our SALT data, can significantly improve the current orbital solutions.

\section*{Acknowledgements}

The authors are grateful to PA Charles for the very valuable discussions. All of the observations reported in this paper were obtained with the SALT under programme 2020-2-MLT-007 (PI: B.\ van Soelen). This work was supported by the Department of Science and Technology and the National Research Foundation of South Africa through a block grant to the South African Gamma-Ray Astronomy Consortium. BvS acknowledges support by the National Research Foundation of South Africa (grant number 119430).
This work made use of \textsc{astropy}:\footnote{http://www.astropy.org} a community-developed core \textsc{python} package and an ecosystem of tools and resources for astronomy \citep{astropy:2013, astropy:2018, astropy:2022}.

\section*{Data Availability}

The data used in this publication are available on reasonable request.



\bibliographystyle{mnras}
\bibliography{references}



\appendix



\bsp	
\label{lastpage}
\end{document}